\documentclass[11pt,twoside]{article}
\usepackage{asp2006, natbib, epsf}

\begin{document}

\title{A New Population of Young Brown Dwarfs}

\author{Kelle L. Cruz,\altaffilmark{1,2} J. Davy Kirkpatrick,\altaffilmark{3}
Adam J. Burgasser,\altaffilmark{4} Dagny Looper,\altaffilmark{5}
Subhanjoy Mohanty,\altaffilmark{6} Lisa Prato,\altaffilmark{7}
Jackie Faherty\altaffilmark{1,8}, and Adam
Solomon\altaffilmark{1,9}}

\altaffiltext{1}{Department of Astrophysics, American Museum of
Natural History, New York, NY, USA.}
\altaffiltext{2}{NSF Astronomy and Astrophysics Postdoctoral
Fellow.}
\altaffiltext{3}{Infrared Processing and Analysis Center,
California Institute of Technology, Pasadena, CA, USA.}
\altaffiltext{4}{Massachusetts Institute of Technology, Kavli
Institute for Astrophysics and Space Research, Cambridge, MA,
USA.}
\altaffiltext{5}{Institute for Astronomy, University of Hawaii, Honolulu, HI, USA.}
\altaffiltext{6}{Harvard-Smithsonian Center for Astrophysics, Cambridge, MA, USA.}
\altaffiltext{7}{Lowell Observatory, Flagstaff, AZ, USA.}
\altaffiltext{8}{Department of Physics and Astronomy, Stony Brook University, Stony Brook, NY, USA.}
\altaffiltext{9}{Department of Astronomy, Yale University, New Haven, CT, USA.}

\begin{abstract}
We report the discovery of a population of late-M and L field
dwarfs with unusual optical and near-infrared spectral features
that we attribute to low gravity---likely uncommonly young,
low-mass brown dwarfs. Many of these new-found young objects have
southerly declinations and distance estimates within 60 parsecs.
Intriguingly, these are the same properties of recently
discovered, nearby, intermediate-age (5--50 Myr), loose
associations such as Tucana/Horologium, the TW Hydrae association,
and the Beta Pictoris moving group. We describe our efforts to
confirm cluster membership and to further investigate this
possible new young population of brown dwarfs.
\end{abstract}

\section{Introduction}

Until recently, studies of brown dwarfs (BD) have largely focused
on two stages of evolution: the very young ($\sim$1~Myr) and the
mature ($\ga$1~Gyr). Brown dwarfs in young clusters are studied
because they are still fairly luminous (typically M type) and the
age of the cluster can be adopted for the brown dwarf. In fact,
most of the posters presented at this conference about brown
dwarfs are studies focused on young clusters: BDs with disks in
Upper Sco (Bouy), low-mass binaries in Taurus (Konopacky), new BDs
in Orion (Faherty), new low-mass stars and BDs in Lupus (Allen),
and new BDs in Trapezium (Riddick). A major drawback of these
clusters is their rather large distance from the Sun
($\sim$100--500~pc) and reliably identifying the lowest-mass
components of these clusters has proven to be a significant
observational challenge.

The advantage of studying brown dwarfs in the field is that they
can be identified in our immediate proximity. Only the
lowest-mass, faintest objects (i.e., L and T dwarfs) within
$\sim$30~pc can be studied in detail. However, only very weak
constraints can be made on the age and mass of L and T dwarfs
identified in the field. In addition, any protoplanetary disk has
long since dissipated and any planetary-mass companions have
cooled beyond current detection limits.

We believe that we have uncovered a nearby ($\sim$30--60~pc),
juvenile (5--50 Myr) population of brown dwarfs that mediate the
current bi-polar situation. The youth of our targets is inferred
from the presence of conspicuous low-gravity features in their
optical and/or near-infrared spectra not seen in hundreds of other
M and L dwarfs; we show two examples in Figure~1. Compared to old
field dwarfs of similar spectral type (equivalent temperature),
low gravity is indicative of both a lower mass and larger
radius---hallmarks of young brown dwarfs still undergoing
gravitational contraction. The spectral features present in our
targets are similar to those seen in members of very young
clusters (e.g., Orion) but since none of our targets are near any
tightly bound groups, they are most certainly older than 1~Myr.
The upper limit on our age estimate is based on the stronger
low-gravity spectral features exhibited in our targets than those
seen in members of 100~Myr old clusters (e.g., Pleiades).
Additionally, as many of our young candidates appear to be likely
members of 8--50~Myr old associations (Figure 2), adopting an age
range of 5--50~Myr for similar objects is reasonable.

The prototype of this new population of low-gravity L dwarfs is
2MASS J01415823$-$\-4633574 (hereafter 2M~0141$-$46). This object
resembles a L0 dwarf but has spectral features indicative of a
very low gravity. The discovery and detailed analysis of this
object is presented by \citet{kc_Kirkpatrick06}. Comparing the
optical and near-infrared spectra of to 2M~0141$-$46 to both
models and dwarfs with known ages, these authors estimate a mass
of 6--25~$M_{Jup}$ and an age 1--50~Myr.

Another benchmark object is 2MASS~J12073346$-$\-3932539
(hereafter\- 2M 1207$-$39). This $\sim$25~$M_{Jup}$ object has a
spectral type of M8, low-gravity spectral features, and is a
confirmed member of the $\sim$10~Myr old TW Hydrae Association
(TWA) \citep{kc_Gizis02, kc_Mohanty03}. \citet{kc_Chauvin04}
detected a $\sim$5--10~$M_{Jup}$ mid-to-late L dwarf companion
(hyped as the first direct detection of an exo-planet) with a
$\sim$0\farcs8 separation from the primary. More recently,
evidence has been found that each component has its own disk
\citep{kc_Riaz06,kc_Mohanty07}. The 2M~1207$-$39 system
demonstrates not only that juvenile brown dwarfs harbor disks, but
that they have a fundamental role to play in the study of star,
brown dwarf, and planet formation. (Also see the poster
contributions by Riaz and Stelzer on the 2M~1207$-$39 disks and
Kasper on low-mass companions to stars in the Tucana and
$\beta$~Pictoris moving groups.)

We have identified $\sim$50 objects similar to 2M~0141$-$46 and
2M~1207$-$39. In \S~2 we present our spectral observations that
indicate low-gravity and youth. The distribution of these objects
on the sky and the similarity to the nearby, young associations
and moving groups such as AB Doradus and Tucana/Horlogium are
discussed in \S~3. In \S~4 we describe our extensive observational
campaign to study this new population. Finally, we summarize our
conclusions in \S~5.

\section{Observations}

We have been using data from the Two Micron All Sky Survey
\cite[2MASS]{kc_2MASS} to undertake extensive, all-sky photometric
searches for nearby late-M and L dwarfs (\citealt{kc_Paper9}; Reid
et al., in prep.; Looper et al., in prep.). As part of these
searches, we have been obtaining both optical (6000--10,000~\AA)
and near-infrared (0.8--2.5~\micron) spectra of candidate nearby L
dwarfs. As mentioned in several papers
(\citealt{kc_Cruz03,kc_Paper9,kc_Kirkpatrick06}; Reid et al., in
prep.; Looper et al., in prep.) this follow-up has uncovered
several objects with optical spectral features indicative of
low-gravity. In our combined efforts, we have uncovered over
$\sim$50 late-M and L dwarfs with low-gravity features in the
optical and/or near-infrared.

\begin{figure}[t]
\plottwo{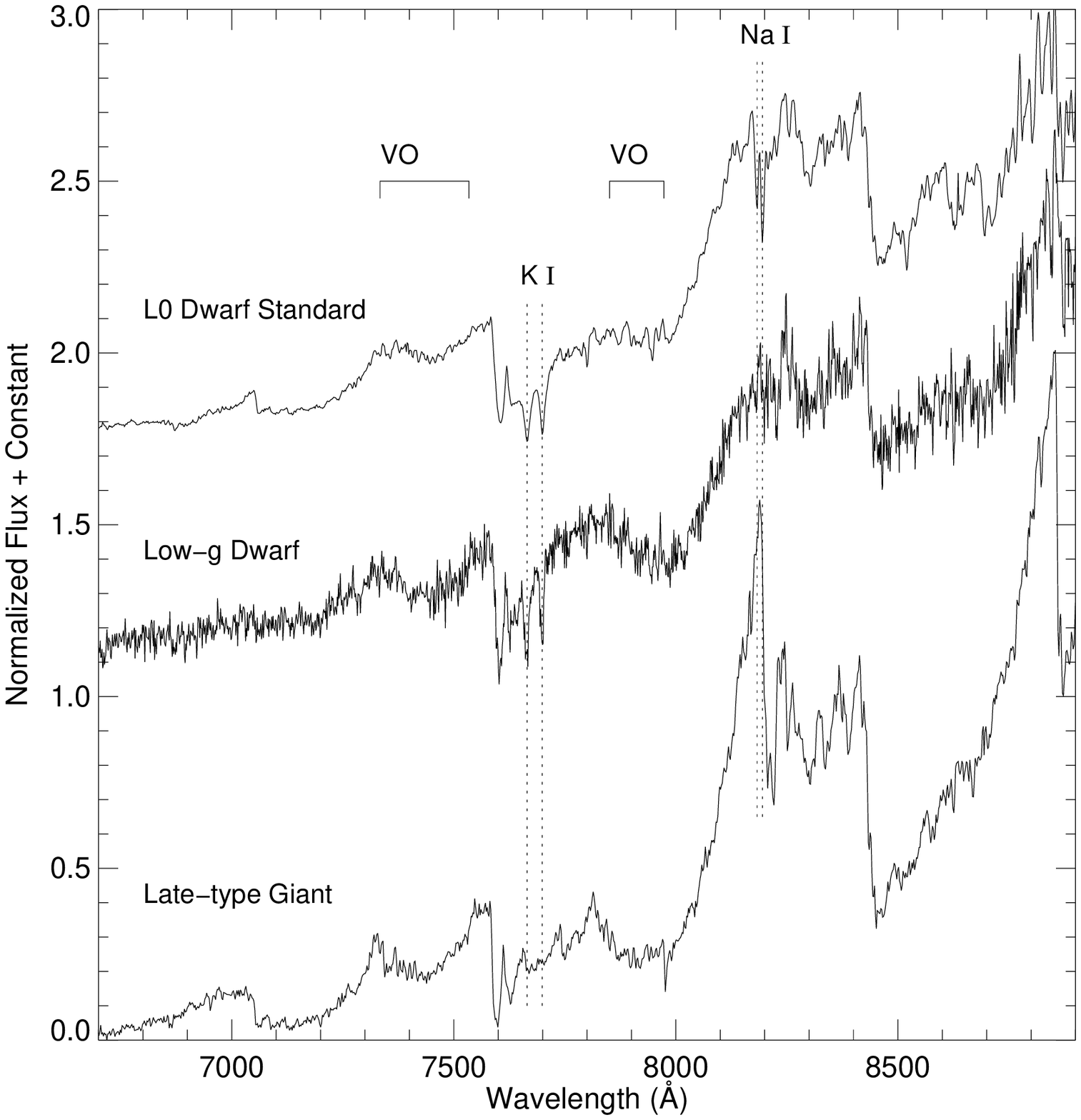}{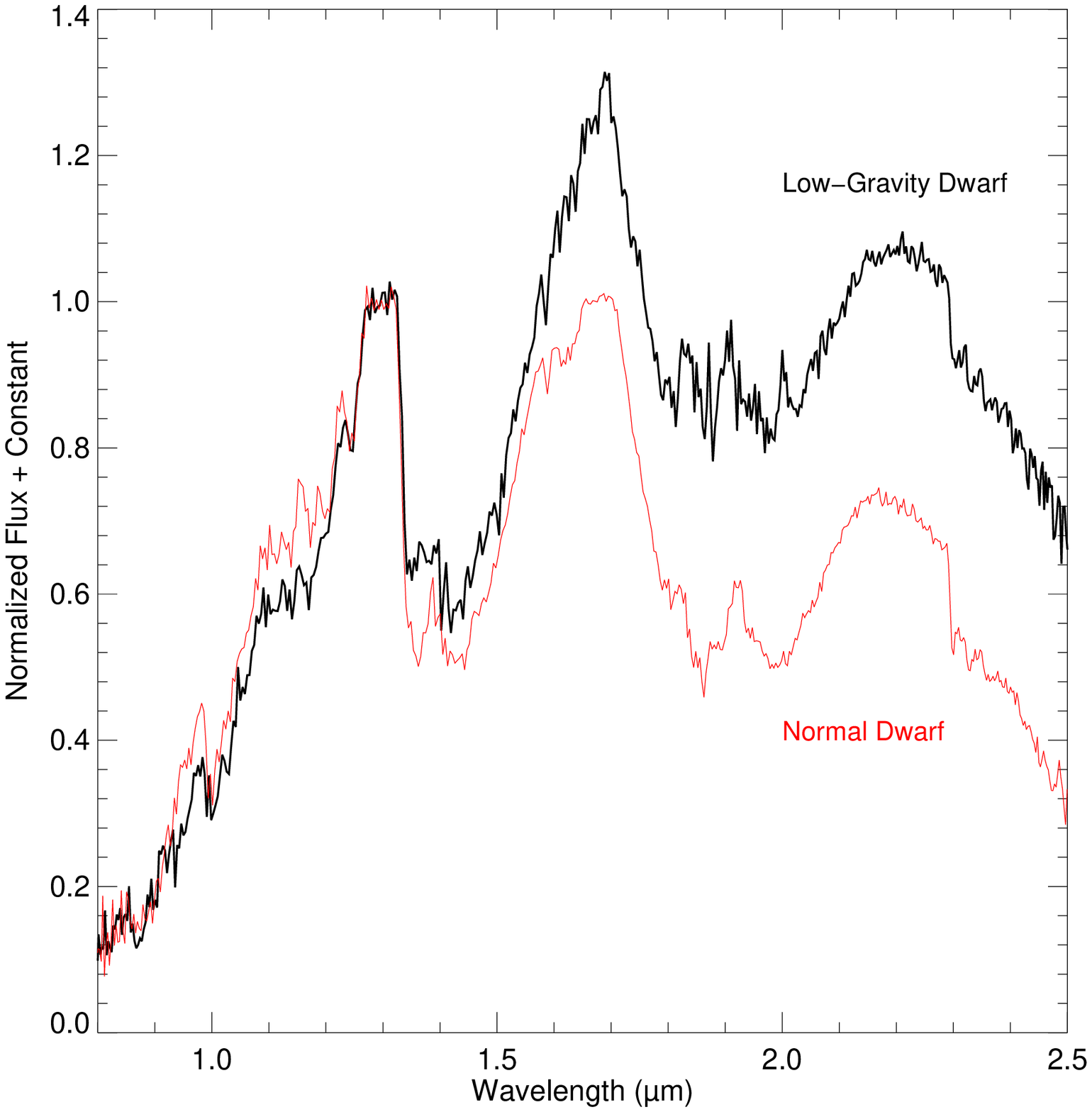}
\caption{\textit{Left}: Optical spectra of a normal L0 dwarf, a
low-gravity dwarf, and a late-type giant (\textit{top-to-bottom}).
The VO, K I, and Na I absorption strengths of the low-gravity
dwarf are between the normal dwarf and the giant. \textit{Right}:
Near-infrared spectra of a normal L5 dwarf (\textit{red/grey}) and
a low-gravity dwarf (\textit{thick black}). The low-gravity object
is characterized by a red spectral slope and a peak-shaped H band
(1.4--1.8~\micron).}
\end{figure}

Our optical data have been obtained with NOAO facilities in both
hemispheres and the Keck I telescope; these observations are
described in detail in \citet{kc_Cruz03} and \citet{kc_K99}. The
left panel of Figure~1 compares the optical spectrum of a
low-gravity L dwarf to spectra of a normal dwarf and a giant.

Near-infrared spectra have been obtained with the SpeX
spectrograph on the NASA Infrared Telescope Facility
\citep{kc_Spex}. We have used the low-resolution (R$\sim$250)
prism and the cross-dispersed higher-resolution (R$\sim$2000)
modes to cover 0.8--2.4~\micron. The right panel of Figure~1
overplots the near-infrared prism spectra of a low-gravity dwarf
and a normal dwarf.

The physical origins of the low-gravity spectral features are
discussed extensively in \citet{kc_Kirkpatrick06} and in
Kirkpatrick's contribution to this volume. In addition, in
\citet{kc_Allers07} and in her poster contribution, the
gravity-sensitive features present in the near-infrared spectra of
young brown dwarfs are described and quantified.

\section{Distribution on the Sky}

Figure 2 shows the location of our candidate young brown dwarfs
(five-pointed stars) with confirmed members of nearby associations
AB Dor, $\beta$ Pic, Tuc/Hor, and TWA as identified by
\citet{kc_Zuckerman04}. The spatial distributions of the two
populations, widely distributed and clumped in the south, are
suggestively similar. This is not too surprising since the age and
distance estimates of our young brown dwarfs are consistent with
those of the moving groups. It is worth noting here that our
searches for nearby late-type dwarfs cover the entire sky and the
only major position cut was for the galactic plane. \citep[For a
detailed discussion of our sky coverage, see][Reid et al., in
prep.]{kc_Paper9}.

Intriguingly, there are several targets that are far removed from
the southern young associations. These objects might be the
evaporated population of lower mass objects from these young
associations, part of some known group that we are unaware of, or
could indicate more new young, nearby associations.

\begin{figure}[t]
\plotone{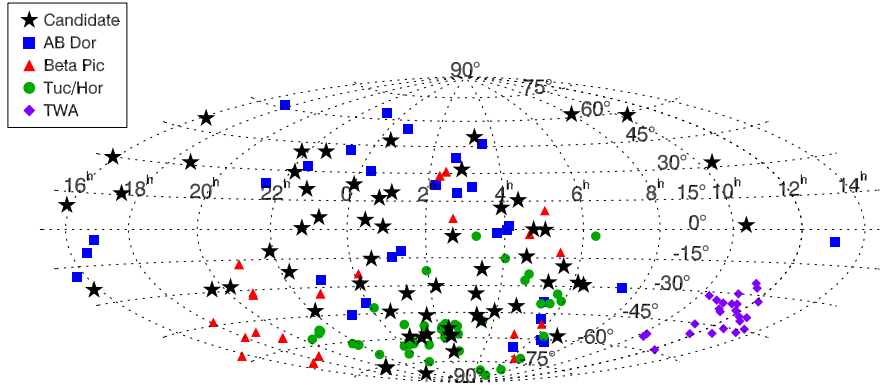}
\caption{Celestial distribution of young brown dwarf candidates
(\textit{five-pointed stars}) on the sky. Also shown are known
members of the AB Dor (\emph{blue squares}), $\beta$ Pic
(\emph{red triangles}), Tuc/Hor (\emph{green circles}), and TWA
(\emph{purple diamonds}) young stellar associations
\citep{kc_Zuckerman04}.}
\end{figure}

\section{Follow-up: Present and Future}

To confirm our suspicion that our candidates are both young and
members of the southerly associations, significant follow-up
observations are being undertaken. A rough kinematic association
can be estimated with just a proper motion. For the brightest
objects, we have used Digital Sky Survey and 2MASS to measure
motions of this sample \citep{kc_Sarah}. Second epoch images of
the fainter targets are being obtained with AMNH SMARTS access to
the CPAPIR wide-field (35 $\times$ 35$\arcmin$) near-infrared
imager on the CTIO 1.5~m telescope.

Radial velocities and trigonometric parallaxes are required to
derive accurate space motions. Radial velocities of our brightest
candidates ($J<15$) are being measured with the optical,
high-resolution spectrograph MIKE on Magellan. We are endeavoring
to use high-resolution, near-infrared spectrographs on 8--10~m
telescopes (e.g., NIRSPEC on Keck) to measure the radial
velocities of our fainter targets ($J>15$). We are currently using
Stony Brook SMARTS access to the ANDICAM near-infrared imager on
the 1.3~m telescope at CTIO to measure the parallaxes (and proper
motions) of our southerly candidates.

Spectral coverage from 0.8--2.5~$\micron$ is needed for all of the
young candidates in order to fully study the low-gravity spectral
features. Currently, not all candidates have spectral coverage in
both the optical and near-infrared and, in some cases, higher
signal-to-noise spectra are needed. We have ongoing programs using
SpeX on the IRTF to cover the near-infrared and we are hoping to
use optical spectrographs on 8--10~m telescopes (e.g., LRIS on
Keck, GMOS on the Gemini telescopes) to obtain optical data.

It is now known that it is not unusual for young brown dwarfs to
harbor disks (see the splinter session summary of Apai \& Luhman,
this volume). Our new-found population of brown dwarfs with older
ages has the potential to lend insight on disk evolution and
planet formation. We are targeting our candidates with Spitzer
IRAC and 24~$\micron$ imaging to investigate the frequency and
properties of brown dwarf disks at intermediate ages.

\section{Summary and Conclusions}

We have uncovered a new population of young, low-mass brown
dwarfs. Based on their spectral appearance and spatial
distribution, we conclude that these objects most likely are
5--50~Myr old and $\sim$10--$30~M_{Jup}$.

This juvenile population will enable the quantification and
calibration of the age and mass-sensitive features present in
brown dwarf spectra. This population also provides an opportunity
to study the epoch of disk evolution when planet formation is
thought to be ongoing. These objects possibly also harbor brown
dwarf companions with masses overlapping the planetary-mass regime
providing a laboratory ideal for identifying observational
differences between brown dwarfs and planets.

\acknowledgements

K. L. C. is supported by an NSF Astronomy and Astrophysics
Postdoctoral Fellowship under AST 04-01418.

\end{document}